\documentclass[12pt]{article}
\usepackage{graphicx}

\newsavebox{\sboxpubnumber}
\newsavebox{\sboxpubdate}
\newcommand{\pubdate}[1]{\begin{lrbox}{\sboxpubdate}{#1}\end{lrbox}}
\newcommand{\pubnumber}[1]{\begin{lrbox}{\sboxpubnumber}{\begin{tabular}{l} #1 \\
                 \usebox{\sboxpubdate}
                 \end{tabular}}
                           \end{lrbox}
                           \pubblock}
\newcommand{\Title}[1]{\begin{center} {\Large #1 } \end{center}}
\newcommand{\Author}[1]{\begin{center}{ \sc #1} \end{center}}
\newcommand{\Address}[1]{\begin{center}{ \it #1} \end{center}}

\newcommand{\pubblock}{\rightline{
            \usebox{\sboxpubnumber}}}
\newenvironment{Abstract}{\begin{quotation}  }{\end{quotation}}
\newenvironment{Presented}{\begin{quotation} \begin{center}
             PRESENTED AT\end{center}\bigskip
      \begin{center}\begin{large}}{\end{large}\end{center}
      \end{quotation}}

\begin{document}

\begin{titlepage}
\pubdate{\today}                    
\pubnumber{SNUTP 01-046} 

\vfill \Title{Self-tuning Solution of Cosmological Constant in
RS-II Model and Goldstone Boson} \vfill \Author{Jihn E. Kim
\footnote{ This work is supported in part by the BK21 program of
Ministry of Education, the KOSEF Sundo Grant, and by the Center
for High Energy Physics(CHEP), Kyungpook National University.}}
\Address{Department of Physics and Center for Theoretical Physics,
Seoul National University, Seoul 151-747, Korea} \vfill
\begin{Abstract}
I give a review on the self-tuning solution\cite{kklcc} of the
cosmological constant in a 5D RS-II model using a three index
antisymmetric tensor field $A_{MNP}$. The three index
antisymmetric tensor field can be the fundamental one appearing in
11D supergravity. Also, the dual of its field strength $H_{MNPQ}$,
being a massless scalar, may be interpreted as a Goldstone boson
of some spontaneously broken global symmetry.
\end{Abstract}
\vfill
\begin{Presented}
    COSMO-01 \\
    Rovaniemi, Finland, \\
    August 29 -- September 4, 2001
\end{Presented}
\vfill
\end{titlepage}
\def\thefootnote{\fnsymbol{footnote}}
\setcounter{footnote}{0}


\section{Introduction}

The cosmological constant problem is the most severe hierarchy
problem or fine-tuning problem known to particle physicists since
1975\cite{veltman,weinberg}.

Another well-known hierarchy problem is the gauge hierarchy
problem encountered in GUT. In GUT's there appear two scales which
differ by a factor of $10^{28}$. At the classical Lagrangian
level, there appear parameters of the GUT scale which is of order
$10^{32}$~GeV$^2$. But the loop corrections and GUT symmetry
breaking must be considered. The known hierarchy requires the
difference must be of order $\sim$TeV$^2$ after including all
these effects, i.e. $M_1^2+M_2^2=O(10^{-28})$~GeV$^2$. To achieve
this small number, we have to fine-tune the parameters $M_1^2$ and
$M_2^2$ both of which appear in the Lagrangian. Supersymmetry has
been employed to understand this gauge hierarchy problem.

Gravity is described by the metric tensor $g_{\mu\nu}$. The Rieman
tensor $R_{\mu\nu}$ is the other second rank tensor. The Einstein
equation is obtained by the variation of the action proportional
to $\int d^4x\sqrt{-g}R$, where $g$ is the determinant of the
metric and $R=g^{\mu\nu} R_{\mu\nu}$. But a general form of the
action invariant under the general coordinate transformation can
be written as
\begin{equation}
S=\int d^4x \sqrt{-g}\left\{ \frac{M^2}{2}R-V_0+\cdots
\right\}\label{EHL}
\end{equation}
where $M^2$ is proportional to the inverse Newton constant $G$
($M^2=8\pi/G$), $V_0$ is a pure constant and the ellipses denote
the other pieces in the Lagrangian whose vacuum expectation
value(VEV) vanishes. The variation of the above action leads to
the gravity equation
\begin{equation}
R_{\mu\nu}-\frac{1}{2}Rg_{\mu\nu}-8\pi GV_0g_{\mu\nu}=8\pi
GT_{\mu\nu}\label{ehleq}
\end{equation}
where the energy momentum tensor $T_{\mu\nu}$ is obtained from the
ellipses. The term $8\pi GV_0$ is the so-called cosmological
constant(c.c.) $\Lambda$. Historically, Einstein introduced the
cosmological constant in 1917 to make the universe static, which
was the belief at that time. But the observation of the Hubble
expansion twelve years later invalidated Einstein's need for the
c.c. As we have seen in Eq. (\ref{EHL}), it is quite natural to
introduce a constant in the action.  Therefore, the c.c. problem
could have been formulated in 1910's, 60 years earlier. The
constant is of order the mass scale in question. The parameter $M$
appearing in Eq. (\ref{EHL}) is the Planck mass $M=2.44\times
10^{18}$~GeV which is astronomically larger than the electroweak
scale. Since gravity introduces a large mass $M$, any other
parameter in gravity is expected to be of that order, which is a
natural expectation. Namely, $V_0$ appearing in (\ref{EHL}) is
expected to be of order $M$. However, the bound on the vacuum
energy was known to be $<(0.01\ {\rm eV})^4$, which implied a
fine-tuning of order $10^{-120}$. Thus, c.c. problem has been
known to be the most serious hierarchy problem.

\begin{figure}[htb]
    \centering
    \includegraphics[height=1.5in]{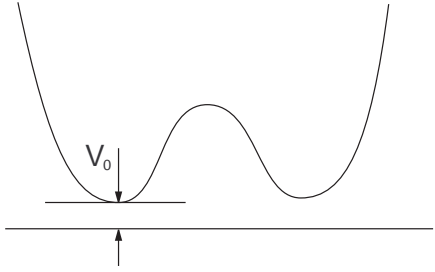}
\caption{The Higgs potential for breaking the $SU(2)\times U(1)$
gauge group.}
  \label{fig:spon}
\end{figure}

Usually, a hierarchy problem is understood if there exists a
symmetry related to it. The difficulty with the c.c. problem is
that there is no such symmetry working. An obvious symmetry one
can imagine is the scale invariance. However it must be badly
broken by the mass terms at the electroweak scale. If the scale
invariance is assumed to be broken at the electroweak scale, we
still have a hierarchy problem of order $10^{56}$.

This c.c. problem surfaced as a very serious one when one
considered the spontaneous symmetry breaking in particle
physics~\cite{veltman}. As shown in Fig. 1, one does not know
whehere to put the minimum point. Except gravity, the position
does not matter. But in gravity its position determines the c.c.
There have been several attempts toward the solution, by
Hawking\cite{hawking}, Witten\cite{witten},
Weinberg\cite{anthropic}, Coleman\cite{coleman}, etc, under the
name of probabilistic interpretation in Euclidian gravity,
boundary of different phases, anthropic solution, wormhole
solution, etc. We will comment more on Hawking's solution later.
The anthropic solution relies on the requirement that life
evolution is not very much affected by the existence of c.c.
However, if c.c. is too large, galaxy formations may be hindered.
Weinberg observed that the requirement for the condensation of
matter needs $\rho_\Lambda<550\rho_c$. Thus, we may need a
fine-tuning but only one part in a thousand.

\section{Self-tuning solutions}

\subsection{Old version}
If there exists a solution for the flat space, it is called a
self-tuning solution, or a solution with the undetermined
integration constant(UIC). In 4D, it is impossible. For a nonzero
$\Lambda$ in 4D, a flat space ansatz $ds^2=d{\bf x}^2-dt^2$ does
not allow a solution. The de Sitter space(dS, $\Lambda>0$) or anti
de Sitter space(AdS, $\Lambda<0$) solution is possible. To reach a
nearly flat space solution, one needs an extreme fine tuning,
which is the c.c. problem.

But suppose that there exists an UIC. Witten used the four index
field strength  $H_{\mu\nu\rho\sigma}$ to obtain an
UIC\cite{witten}. The equation of motion of $H$ leads to an UIC,
say $c$. Thus, the vacuum energy contains a piece $\sim c^2$. This
UIC $c$ can be adjusted so that the final c.c. is zero. Once $c$
is determined, there is no more UIC because $H_{\mu\nu\rho\sigma}$
is not a dynamical field in 4D. When vacuum energy is added later,
there is no handle to adjust further. In a sense, it was another
way of fine-tuning. However, if there exists a dynamical field
allowing an UIC, it is a desired old style self-tuning solution.
This old version did not care whether there also exist de Sitter
or anti de Sitter solutions. Selection of the flat space out of
these solutions is from a principle such as Hawking's
probabilistic choice.

\subsection{New version}
In recent years, a more ambitious attempt was proposed, where only
the flat space ansatz has the solution\cite{kachru}. This idea
attracted a great deal of attention in the Randall-Sundrum II type
models\cite{rs2}. The RS type models were constructed in 5D anti
de Sitter space, i.e. the 5D bulk cosmological constant
$\Lambda_b<0$, with brane(s) located at fixed point(s). The RS II
model uses only one brane. At this brane one can introduce a brane
tension $\Lambda_1$. Thus, the gravity Lagrangian contains two
free parameters $\Lambda_b$ and $\Lambda_1$.

\begin{figure}[htb]
    \centering
    \includegraphics[height=1.5in]{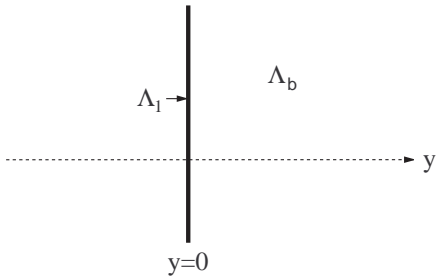}
\caption{The RS II model with a brane at $y=0$.}
  \label{fig:rsII}
\end{figure}
In Fig. 2, we show
this situation schematically, where the extra dimension is the
$y$-direction. The 4D is denoted as $x^\mu$. The 4D flat space
ansatz allows a solution for a specific choice of $k_1$(basically
$\Lambda_1$, $k_1=\Lambda_1/6M^3$) and $k$(basically $\Lambda_b$,
$k=\sqrt{-\Lambda_b/6M^3}$): $k_1=k$. Therefore, it requires a
fine-tuning between parameters as in the 4D case. However, it
seems to be an improvement since we reach at the 4D flat space
from nonzero cosmological constants and the RS II type models
seems to be a good play ground to obtain UIC solutions.

The RS II model is an interesting extension of the space-time
without compactification. With the bulk AdS, the uncompactified
fifth dimension can be acceptable due to the localized gravity.
For the brane located at $y=0$, the action is
\begin{equation}
S=\int d^4x \int dy \sqrt{-g}\left\{
\frac{M^3}{2}(R-\Lambda_b)+(-\Lambda_1+{\cal L}_{matter})
\delta(y) \right\}
\end{equation}
where $M$ is the 5D fundamental scale and ${\cal L}_{matter}$ is
the matter Lagrangian, assuming the matter is located at the brane
only. The flat space ansatz,
\begin{equation}
ds^2=\beta(y)^2\eta_{\mu\nu}dx^\mu dx^\nu+dy^2\label{flat}
\end{equation}
allows the solution if $k=k_1$. Even though the 4D is flat, it is
curved in the direction of the fifth dimension, denoted by the
warp factor $\beta(y)=\beta_0\exp(-k|y|)$. Namely, the gravity is
exponentially unimportant if one is far away from the brane. [If
there were more branes, there are more conditions to satisfy
toward a flat space solution since one can introduce a brane tension at
each brane. Thus, the RS II type models are the simplest ones.]

The try to obtain a new type of self-tuning solution was initiated
by Arkani-Hamed et. al. and Kachru et. al.\cite{kachru}. For
example the 5D Lagrangian with the introduction of a massless bulk
scalar field $\phi$, coupling to the brane tension,
\begin{equation}
{\cal L}=R-\Lambda e^{a\phi}-\frac{4}{3}(\nabla\phi)^2-Ve^{b\phi}
\delta(y)\label{kachru}
\end{equation}
where we set $M^3/2=1$. We may ask, $\lq\lq$Why this Lagrangian?",
which involves more difficult related questions. Accepting this,
we must satisfy the following Einstein and field equations, with
the flat ansatz (\ref{flat}),
\begin{eqnarray}
({\rm dilaton})&:&
\frac{8}{3}\phi^{\prime\prime}+\frac{32}{3}A'\phi'-a\Lambda
e^{a\phi} -bV\delta(y)e^{b\phi}=0 \nonumber\\
(55)&:& 6(A')^2-\frac{2}{3}(\phi')^2+\frac{1}{2}\Lambda e^{a\phi}=0
  \nonumber\\
(55),(\mu\nu)&:&
3A^{\prime\prime}+\frac{4}{3}(\phi')^2+\frac{1}{2}e^{b\phi}V\delta(y)=0
\end{eqnarray}
where $2A(y)=\ln\beta(y)$, and prime denotes the derivative with
respect to $y$. For $\Lambda=0$, there exists a bulk solution
satisfying $A'=\alpha\phi'$,
\begin{equation}
\phi=\pm\frac{3}{4}\ln\left|\frac{4}{3}y+c\right|+d,\
\ \alpha=\pm\frac{1}{3}
\end{equation}
where $c$ and $d$ are determined without fine-tuning of the
parameters. The solution has a singularity at $y_c\equiv -(3/4)c$
or diverges logarithmically at large $|y|$. The logarithmically
diverging solution does not realize the localization of gravity.
If we restrict the space up to the singular point $y_c$, then at
every $y$ inside the space it is flat. However, the effective 4D
theory is the one after integrating out the allowed $y$ space.
Since $y_c$ is the naked singularity, we do not know how to cut
the $y$ integration near $y_c$, implying a possibility that the
flat space ansatz does not lead to a solution. Depending on how to
cut the integral, one may introduce a nonzero c.c. F$\ddot{\rm
o}$rste et. al. tried to understand this problem by curing the
singularity by putting a brane at $y_c$\cite{nilles}. Then, a flat
4D space solution is possible but one needs a fine-tuning. It is
easy to understand. If one more brane is introduced, then there is
one more tension parameter $\Lambda_2$, i.e. in the Lagrangian one
adds $\Lambda_2\delta(y-y_c)$. If the space is flat for one
specific value of $\Lambda_2$, then it must be curved for the
other values of $\Lambda_2$, since the $y$ integration gives a
c.c. contribution directly from $\Lambda_2$.

This example teaches us that the self-tuning solution better
should not have a singular point in the whole $y$ space.

\section{The self-tuning solution with $1/H^2$}

As we have seen in the RS II model, the Einstein-Hilbert action
alone does not produce a self-tuning solution. Inclusion of higher
order gravity does not improve this situation\cite{kklgb}. We need
matter field(s) in the bulk. The first try is a massless spin-0
field in the bulk as Ref. (\cite{kachru}) tried so that it affects
the whole region of the bulk. However, it may be better if there
appears a symmetry in the spin-0 sector. These are achieved by a
three index antisymmetric tensor field $A_{MNP}$. In 5D the dual
of its field strength is interpreted as a scalar. The field
strength $H_{MNPQ}$ is invariant under the gauge transformation
$A_{MNP}\rightarrow A_{MNP}+\partial [M\lambda_{NP}]$, thus
masslessness arises from the symmetry. There will be one $U(1)$
gauge symmetry remaining with one massless pseudoscalar field
which is $a$, $\partial_M a\propto \sqrt{-g}
\epsilon_{MNPQR}H^{NPQR}$. But the
interactions are important for the solution, as Ref.
(\cite{kachru}) find a bulk solution for the specific form of the
interaction.

The first guess is the bulk term $-(M/48)H^2$ where $H^2\equiv
H_{MNPQ}H^{MNPQ}$. The brane with tension $\Lambda_1$ is located
at $y=0$, and the bulk c.c. is $\Lambda_b$. The ansatze for the
solution are
\begin{equation}
{\rm Ansatz\ 1}\ :\ ds^2=\beta(y)^2\eta_{\mu\nu}dx^\mu
dx^\nu+dy^2,\ \ {\rm Ansatz\ 2}\ :\ H_{\mu\nu\rho\sigma}
=\epsilon_{\mu\nu\rho\sigma}\frac{\sqrt{-g}}{n(y)}\label{ansatz0}
\end{equation}
where $\mu,\cdots$ are the 4D indices, and $n(y)$ is a function of
$y$ to be determined. It is sufficient to consider (55) and
($\mu\nu$) components Einstein equations and the $H$ field
equation. By setting $M=1$, we obtain the bulk solution
\begin{eqnarray}
\Lambda_b<0\ &:&\ \beta(|y|)=(\frac{a}{k})^{1/4}[\pm\sinh(4k|y|
+c)]^{1/4} \nonumber\\
\Lambda_b>0\ &:&\
\beta(|y|)=(\frac{a}{k})^{1/4}[\sin(4k|y|+c^\prime)
]^{1/4}  \nonumber\\
\Lambda_b=0\ &:&\ \beta(|y|)=|4a|y|+c^{\prime\prime}|^{1/4}.
\end{eqnarray}
For a localizable (near $y=0$) metric, there exists a singularity
at $y=-c/4k$, etc., except for some cases with $\Lambda_b>0$.
Thus, for these singular cases another brane is necessary to cure
the singularity, and we need a fine-tuning as in the case of
Kachru et. al.\cite{kachru}. The bulk de Sitter space without a
singularity is worth commenting. Such a solution is periodic and
depicted in Fig. 3.
We can consider only $|y|\le y_c$, then $\beta^\prime=0$
at $y=\pm y_c$. The boundary condition at $y=0$ determines $c'=
\cot^{-1}(k_1/k)$ and the boundary condition at $y_c$ determines
$y_c$ such that $c'=4ky_c-\cot^{-1}(k_2/k)$, so it looks like an
UIC. But for $y_c$ to behave like an undetermined integration
constant, it should not appear in the equations of motion. Note,
however, that $y_c$ is the VEV of the radion $g_{55}$, and hence
it cannot be a strictly massless Goldstone boson. If it were
massless, it will serve to the long range gravitational
interaction and hence give different results from the general
relativity predictions in the light bending experiments.
Therefore, it should obtain a mass and hence $y_c$ is not a free
parameter but fixed. So the boundary condition at $y_c$ is a
fine-tuning condition\cite{radion}.

\begin{figure}[htb]
    \centering
    \includegraphics[height=2.5in]{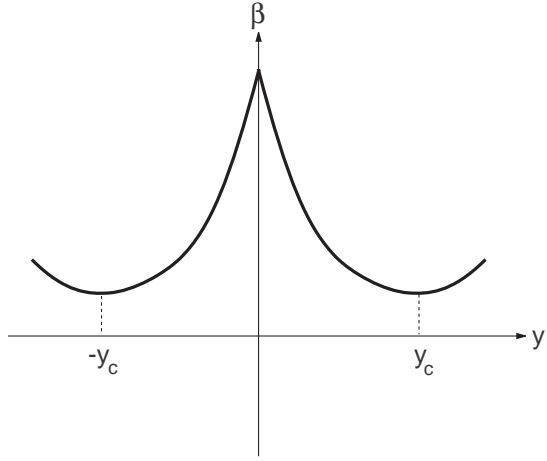}
\caption{The flat space solution with $H^2$.}
  \label{fig:h2}
\end{figure}

In the remainder of this talk, I present a working self-tuning
model. Let us consider the $1/H^2$ term,
\begin{equation}
S=\int d^4x\int dy\sqrt{-g}\left( \frac{1}{2}R +\frac{2\cdot
4!}{H^2}-\Lambda_b-\Lambda_1\delta(y) \right).\label{action}
\end{equation}

\subsection{Flat space solution}

For the flat space ansatze, we use
\begin{equation}
ds^2=\beta(y)^2\eta_{\mu\nu}dx^\mu dx^\nu +dy^2,\ \
H_{\mu\nu\rho\sigma}=\epsilon_{\mu\nu\rho\sigma}\frac{\sqrt{-g}}{n(y)},
\ \ H_{5\mu\nu\rho}=0.\label{ansatz1}
\end{equation}
The $H$ field equation is
$\partial_M[\sqrt{-g}H^{MNPQ}/H^4]=\partial_\mu [\sqrt{-g}H^{\mu
NPQ}/H^4]=0$, and hence fixes $n$ as a function of $y$ only. The
two relevant Einstein equations are
\begin{eqnarray}
(55)\ &:&\
6\left(\frac{\beta^\prime}{\beta}\right)^2=-\Lambda_b-\frac{\beta^8}{A}
\nonumber\\
(\mu\nu)\ &:&\ 3\left(\frac{\beta^\prime}{\beta}\right)^2
+3\left(\frac{\beta^{\prime\prime}}{\beta}\right)= -\Lambda_b
-\Lambda_1\delta(y)-3\left(\frac{\beta^8}{A}\right)
\end{eqnarray}
where $\Lambda_b<0$ and $2n^2=\beta^8/A$ with $A>0$. We require
the $Z_2$ symmetry, and the bulk equation is easily solved. The
boundary condition at $y=0$ is
$\beta^\prime/\beta|_{0^+}=-\Lambda_1/6$. Then, we find a solution
\begin{equation}
\beta(|y|)=\frac{(k/a)^{1/4}}{\left[\cosh(4k|y|+c)\right]^{1/4}}
\label{solution}
\end{equation}
where
\begin{equation}
k=\sqrt{\frac{-\Lambda_b}{6}},\ \ a=\sqrt{\frac{1}{6A}},\ \
k_1=\frac{\Lambda_1}{6}.
\end{equation}
The flat solution is shown in Fig. 4.

\begin{figure}[htb]
    \centering
    \includegraphics[height=2.5in]{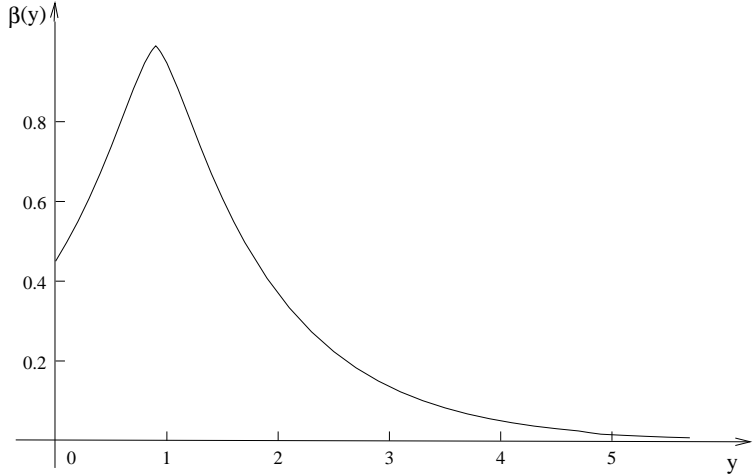}
\caption{The flat space solution.}
\label{fig:beta0}
\end{figure}

This solution has the integration constants $a$ and $c$. $a$
is basically the charge of the universe and determines the 4D
Planck mass. $c$ is the UIC which is fixed by the boundary
condition at $y=0$,
\begin{equation}
\tanh(c)
=\frac{k_1}{k}=\frac{\Lambda_1}{\sqrt{-6\Lambda_b}}.\label{uic}
\end{equation}
This solution shows that, for any value of $\Lambda_1$ in the
finite region allowing $\tanh$, it is possible to have a flat
space solution. Even if the observable sector adds some constant to
$\Lambda_1$, still it is possible to have the flat space solution,
just by changing the shape little bit via $c$. The change is
acceptable since $H$ is a dynamical field. Note that $\beta(y)$ is
a decreasing function of $|y|$ in the large $|y|$ region and it
goes to zero exponentially as $|y|\rightarrow\infty$. This
property is needed for a self-tuning solution.

The key points found in our solution are\\
{\bf (i) $\beta$ has no singularity:} Our solution extends to
infinity without singularity, and $\beta^\prime\rightarrow 0$ as
$y\rightarrow\infty$.\\
{\bf (ii) 4D Planck mass is finite:} Even if the extra dimension
is not compact, this theory can describe an effective 4D theory
since gravity is localized. Integrating with respect to $y$, we
obtain an effective 4D Planck mass which is finite
\begin{equation}
M^2_{4D\ Planck}=\int_{-infty}^\infty dy \beta^2=
2M^3\sqrt{\frac{k}{a}}\int_0^\infty \frac{1}{[\cosh(4ky+c)]^{1/2}}dy
\end{equation}
and is expected to be of order the fundamental parameters.\\
{\bf (iii) Self-tuning:} We obtained a self-tuning solution. To
check that the 4D c.c. is zero we integrate out the solution. For
that we have to include the surface term also
\begin{equation}
S_{surface}=\int d^4x dy 4\cdot
4!\partial_M[\sqrt{-g}\frac{H^{MNPQ}A_{NPQ}}{H^4}].
\end{equation}
Then the action is
\begin{equation}
S=\int d^4x dy\sqrt{-\eta}\beta^4\left[\frac{1}{2\beta^2}R_4
-4\frac{\beta^{\prime\prime}}{\beta}-6
\left(\frac{\beta^\prime}{\beta}\right)^2-\Lambda_b+\frac{2\cdot
4! }{H^2}-\Lambda_1\delta(y)\right]+S_{surface}.
\end{equation}
Then, $-\Lambda_{4D}$ is the $y$ integral except the $R_4$ term.
One can show that $\Lambda_{4D}=0$ and it is consistent with the
original ansatz of the flat space\cite{kklcc}.

\subsection{De Sitter and anti de Sitter space solutions}

For the de Sitter and anti de Sitter space solutions, the metric
is assumed as
\begin{eqnarray}
&ds^2=\beta(y)^2 g_{\mu\nu}dx^\mu dx^\nu +dy^2\nonumber\\
&g_{\mu\nu}=diag.\left(-1,e^{2\sqrt{\lambda}t},e^{2\sqrt{\lambda}t},
e^{2\sqrt{\lambda}t}\right),\ \ (dS_4\rm\ background\ with\
\lambda>0)\nonumber\\
&g_{\mu\nu}=diag.\left(-e^{2\sqrt{-\lambda}x_3},e^{2\sqrt{-\lambda}x_3},
e^{2\sqrt{-\lambda}x_3},1\right),\ \ (AdS_4\rm\ background\ with\
\lambda<0).
\end{eqnarray}
Note that $k=\sqrt{-\Lambda_b/6},k_1=\Lambda_1/6$, and the 4D
Riemann tensor is $R_{\mu\nu}=3\lambda g_{\mu\nu}$. The (55) and
(00) components equations are
\begin{eqnarray}
&6\left(\frac{\beta^\prime}{\beta}\right)^2-6\lambda\frac{1}{\beta^{2}}
=-\Lambda_b-3\frac{\beta^8}{A}\nonumber\\
&3\left(\frac{\beta^\prime}{\beta}\right)^2+3
\frac{\beta^{\prime\prime}}{\beta}-3\lambda\frac{1}{\beta^2}
=-\Lambda_b-\Lambda_1\delta(y)-3\frac{\beta^8}{A}.
\end{eqnarray}
The 4D c.c. obtained from the above ansatze is $\lambda$. Since we
cannot obtain the solution in closed forms, we cannot show this by
integration. However, we have checked this kind of
behavior\cite{kklcc} in the RS II model, using the Karch-Randall
form\cite{kr}. Here, we show just that the de Sitter and anti de
Sitter space solutions exist, and show the warp factor
numerically. In our model, the $y$ derivative of the metric is
\begin{equation}
\beta^\prime =\pm (k_\lambda^2+k^2\beta^2-a^2\beta^{10})^{1/2},\ \
k_\lambda=(\lambda)^{1/2}
\end{equation}
At $y=y_h$ where $\beta(y_h)=0$, $\beta^\prime(y_h)$ needs not be
zero due to the presence of the nonvanishing $k_\lambda$.
Therefore, there exists a point $y_h$ where $\beta^\prime$ is
finite. It is the de Sitter space horizon. It takes an infinite
amount of time to reach $y_h$. Also, we can see that it is
possible $\beta$ can be nonzero where $\beta^\prime$ is zero. It
is the anti de Sitter space solution. These Anti de Sitter space and
de Sitter space solutions are shown in Figs. 5 and 6.
For the de Sitter space, we can integrate from
$-y_h$ to $+y_h$. As in the Karch-Randall example, it should give
the 4D c.c. $\lambda$. The AdS solution does not give a localized
gravity.
\begin{figure}[htb]
    \centering
    \includegraphics[height=1.5in]{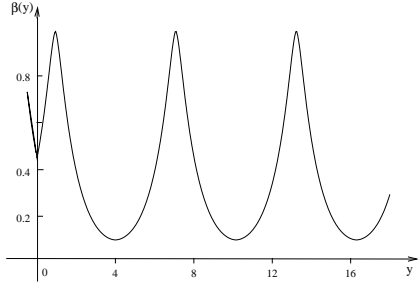}
\caption{The anti de Sitter space solution.} \label{fig:anti}
\end{figure}
\begin{figure}[htb]
\centering
\includegraphics[height=1.5in]{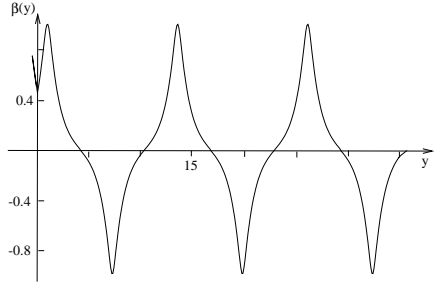}
\caption{The de Sitter space solution.}
\label{desitter}
\end{figure}

In the presence of de Sitter and anti de Sitter space solutions,
the c.c. problem relies on the old self-tuning solution. Namely,
the c.c. is probably zero, following Hawking\cite{hawking}.
Hawking showed that in 4D, with the Euclidian space action
\begin{equation}
-S_E=-\frac{1}{2}\int
d^4x\sqrt{+g}[(R+2\Lambda)+(1/48)H_{\mu\nu\rho\sigma}
H^{\mu\nu\rho\sigma}]
\end{equation}
where we use the unit $8\pi G=1$. The Einstein equation and field
equations are
\begin{eqnarray}
&R_{\mu\nu}-(1/2)g_{\mu\nu}R=\Lambda
g_{\mu\nu}-T_{\mu\nu}\nonumber\\
&\partial_\mu[\sqrt{g}H^{\mu\nu\rho\sigma}]=0.\nonumber
\end{eqnarray} From the field equation, one has
$H^{\mu\nu\rho\sigma}=(1/\sqrt{g})\epsilon^{\mu\nu\rho\sigma} c$
or $H^2=4!c^2$. Thus, he obtained $T^{\mu\nu}=-\Lambda_H
g^{\mu\nu}, \Lambda_H=-c^2/2, R=-4\Lambda_{eff},
\Lambda_{eff}=\Lambda+\Lambda_H$. Thus,
\begin{equation}
-S_E=-(1/2)\int d^4x\sqrt{g}(R+2\Lambda_{eff})={\rm Volume}\cdot
(\Lambda+\Lambda_H)=\frac{3M^4}{\Lambda_{eff}}
\end{equation}
which is maximum at $\Lambda_{eff}=0^+$ which is shown in Fig. 5.
\begin{figure}[htb]
\centering
\includegraphics[height=1.5in]{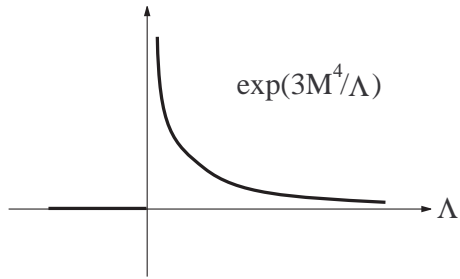}
\caption{Hawking's probability}
  \label{fig:hawking}
\end{figure}
Note that Hawking used equations of motion.
Duff, on the other hand, used the action itself to calculate the
Euclidian action, and obtained
$-S_E=3M^4(\Lambda-(3/2)c^2)/(\Lambda-(1/2)c^2)^2$ which is
minimum at $0^+$\cite{duff}. But the consideration of the surface
term in the action would give additional contribution and should
give Hawking's result. The surface term is essential as we have
shown in our self-tuning solution.

Thus, the maximum probability occurs when the c.c. is zero. Our
self-tuning solution relies on this probabilistic choice of the
flat one from the flat, de Sitter and anti de Sitter space
solutions.

\section{Goldstone boson scenario?}

One point to consider is whether the peculiar kinetic energy term
can be made sensible. Actually we can construct an example which
has reasonable terms. Let us introduce a $U(1)$ gauge field
strength $F_{MN}$ and its coupling to $H_{MNPQ}$ as
\begin{equation}
-\frac{1}{8}H^2(F^2)^2-\frac{1}{4}F^2.
\end{equation}
The Gaussian integral of $A_M$ would choose $F^2=-1/H^2$, and we
would obtain the desired $1/H^2$ term. Therefore, consideration of
$1/H^2$ can be meaningful. But the question is why there is no
term with $H^2$ in the first place.

Since $H$ is a massless boson, it can be considered as a Goldstone
boson. Thus, one may try to construct a theory where a
pseudoscalar Goldstone boson is a kind of cosmion, self-tuning the
c.c. at zero. The question is how one obtains the term $1/H^2$
instead of $H^2$.


\end{document}